\documentclass[aps,pre,twocolumn,10pt,prx]{revtex4-1}

\usepackage[latin1]{inputenc} 
\usepackage[T1]{fontenc}
\usepackage{lmodern}
\usepackage{graphicx}
\usepackage{amsmath,amssymb,color,bm}

\newcommand{\LBB}[1]{{\color{black} { #1}}}
\newcommand{\LC}[1]{{\color{black} {#1}}}
\newcommand{\LB}[1]{{\color{black} {#1}}}
\newcommand{\degree}{$^\circ$}

\begin{document}
\title{Nanorheology of interfacial water during ice gliding}

\author{L. Canale$^1$, J. Comtet$^1$, A. Nigu\`es$^1$, C. Cohen$^2$, C. Clanet$^2$, A. Siria$^{1,\star}$, L.Bocquet$^{1,\star}$}
\affiliation{$^{1}$Laboratoire de Physique de l'Ecole normale Sup\'erieure, ENS, Universit\'e PSL, CNRS, \\
Sorbonne Universit\'e, Universit\'e Paris-Diderot, Sorbonne Paris Cit\'e, Paris, France, \\
$^{2}$LadHyX, UMR CNRS 7646 , Ecole Polytechnique, 91128 Palaiseau, France}
\email{lyderic.bocquet@lps.ens.fr, alessandro.siria@lps.ens.fr}

\begin{abstract}
The slipperiness of ice is an everyday-life phenomenon which, surprisingly, remains controversial despite a long scientific history. 
The very small friction measured on ice is classically attributed to the presence of a thin self-lubricating  film of meltwater between the slider and the ice. But  while the macroscale friction behavior of ice and snow has been widely investigated, very little is known about the interfacial water film and its mechanical properties. In this work, we develop a stroke-probe force measurement technique to uncover the microscopic mechanisms underlying ice lubrication. We simultaneously measure the shear friction of a bead on ice and quantify the in-situ mechanical properties of the interfacial film, as well as its thickness, under various regimes of speed and temperature. 
In contrast with standard views, meltwater is found to exhibit a complex visco-elastic rheology, with a  viscosity up to two orders of magnitude larger than pristine water. The non-conventional rheology of meltwater provides a new, consistent, rationale for ice slipperiness. 
Hydrophobic coatings are furthermore shown to strongly reduce friction due to a surprising change in the local viscosity, providing an unexpected explanation for waxing effects in winter sports. Beyond ice friction, our results suggest new avenues towards self-healing lubricants to achieve ultra-low friction. 
\end{abstract}

\maketitle

\section{Introduction}

Ice and snow exhibit outstanding friction properties, with exceptionally low friction coefficients \cite{Persson}.
This unique behavior is at the root of all winter sports \cite{Rosenberg2005,Oosterkamp2019}; it is also a major source of risk in transportation,
or, in a very different context, a key ingredient in glacier sliding \cite{Paterson2016,Persson2018}.
The low friction of ice remains however highly counterintuive and paradoxical, 
since, comparatively, liquid water is usually considered as a bad lubricant due to its low viscosity.
In spite of more than a century of investigation, the very origin of this puzzling property is not settled yet and {remains} highly debated.
Since the seminal work of 
Faraday \cite {Faraday1859}, a consensus has been reached on the existence of a liquid-like layer wetting the ice surface \cite{Doppenschmidt2000,Salmeron2002,Kornienko2013,Wettlaufer1999,Conde2008,Bonn2019}, with a  thickness varying between 1 and 100 nm depending on the temperature \cite{Michaelides2017}, 
although the underlying mechanism of formation remains debated \cite{Sanchez2017,Murata2016}.
Now, under sliding the fate of the interfacial film remains largely {unknown} \cite{Persson2015}. 
The pioneering works of Bowden \cite{Bowden1939,Bowden1955} and later Colbeck \cite{Colbeck1988} have discarded pressure-melting mechanisms in ice and snow friction and suggested frictional-melting: viscous dissipation generates heat which rises the temperature in the contact region up to the melting point, hereby creating a water lubricating film. This scenario has been further explored by numerous {\it macroscopic} tribological measurements \cite{Buhl2001,Baurle2006,Kietzig2009,Kietzig2010},  supported -- at least partly -- by theoretical frameworks \cite{Oosterkamp2019,Colbeck1988,Evans1976,Persson2015}.
However, probing the {\it in-situ} properties of this interfacial film remains a formidable challenge. \LB{Indeed the meltwater film is dynamically and self-consistently generated under sliding, which makes the ice-water boundary elusive to detect.} Also, standard interferometry techniques fail because of the low contrast of the interface. The few existing measurements focused on the meltwater film thickness and led to contradictory results, with the estimated thickness varying from 5-10 $\mu$m \cite{Ambach1981} to less than 50 nm \cite{Strausky1998}. {Even} more puzzling, recent local temperature measurements have precluded full melting of an interfacial water film under sliding, contradicting standard explanations \cite{Lever2018}. 
It is not an understatement that the fundamental mechanisms for the slipperiness of ice (and snow) still remain a mystery.
In this work we propose a radically new approach which enables to disentangle the various physical ingredients at stake. We investigate simultaneously the friction of a millimetric slider on ice, and the corresponding interfacial mechanical properties of the meltwater film at the nanoscale.
To this end, we harvest the possibilities offered by a newly introduced force measurement apparatus \cite{Canale2018,Laine2019}, to realize here a ``stroke-probe'' tribometer with nanoscopic sensitivity. This apparatus allows us to close the gap between nano- and macro- scale tribometry  \cite{Carpick2019}, which is a prerequisite for the investigation of the ice interface under sliding. 

\section{Stroke-probe AFM experiments}

The experimental setup, shown in Fig.~1A, \LB{consists of a double-mode Tuning Fork Atomic Force Microscope (TF-AFM)  \cite{Canale2018,Laine2019,Comtet2017}.} The setup is placed in a cold chamber with controlled temperature from -16 to 0 $^{\circ}$C and 70-80 \% relative humidity. \LB{A centimetric sample of ice is obtained from deionized water . Under the present conditions, ice evaporation was measured to occur on a timescale much longer than the measurements}. The ice temperature is directly monitored via an embedded thermocouple. A millimetric borosilicate glass bead  (Fig.~1A;  Fig.~S2 of the Supplemental Materials \cite{SM} for bead characterization) is glued on one prong of a centimetric aluminium tuning fork. The system can be accurately modeled as a mass-spring resonator of large stiffness $K_\text{T} \approx 10^2$ kN.m$^{-1}$ and quality factor $Q_\text{T} \approx 2500$. An electromagnetic excitation at the tuning fork resonance frequency $f_\text{T}\simeq 560 \text{ Hz}$ 
then yields a lateral oscillatory motion of the bead parallel to the ice surface, Fig.~1A (red arrow). Its amplitude $a_\text{T}$ and phase shift $\phi_\text{T}$  with respect to the excitation force are monitored with an accelerometer glued on one prong. The oscillating sphere is brought into contact with the ice surface by a piezo element with integrated position sensor of nanometric resolution. The  lateral stroke of the sphere then shears the ice with an amplitude $a_\text{T}\sim 1-30\,\mu$m and velocity  $U = {2 \pi a_\text{T} f_\text{T}}$, typically up to $0.1$ m.s$^{-1}$.
A Phase-Lock-Loop (PLL) maintains the system at the resonance by tuning the excitation frequency $f_\text{T}$ and the tangential friction force $F_\text{F}$ is simply measured by tracking the excitation force $F_\text{T}^{em}$ necessary to keep the oscillation amplitude $a_\text{T}$ constant while sliding 
according to $F_{F}=\frac{K_\text{T}}{Q_\text{T}} \left( \frac{F_\text{T}^{em}}{F_\text{T,0}^{em}}-1 \right) \cdot a_\text{T}$ \cite{Laine2019}.
\begin{figure}[!htb]
\centering
  \includegraphics[width=1\columnwidth]{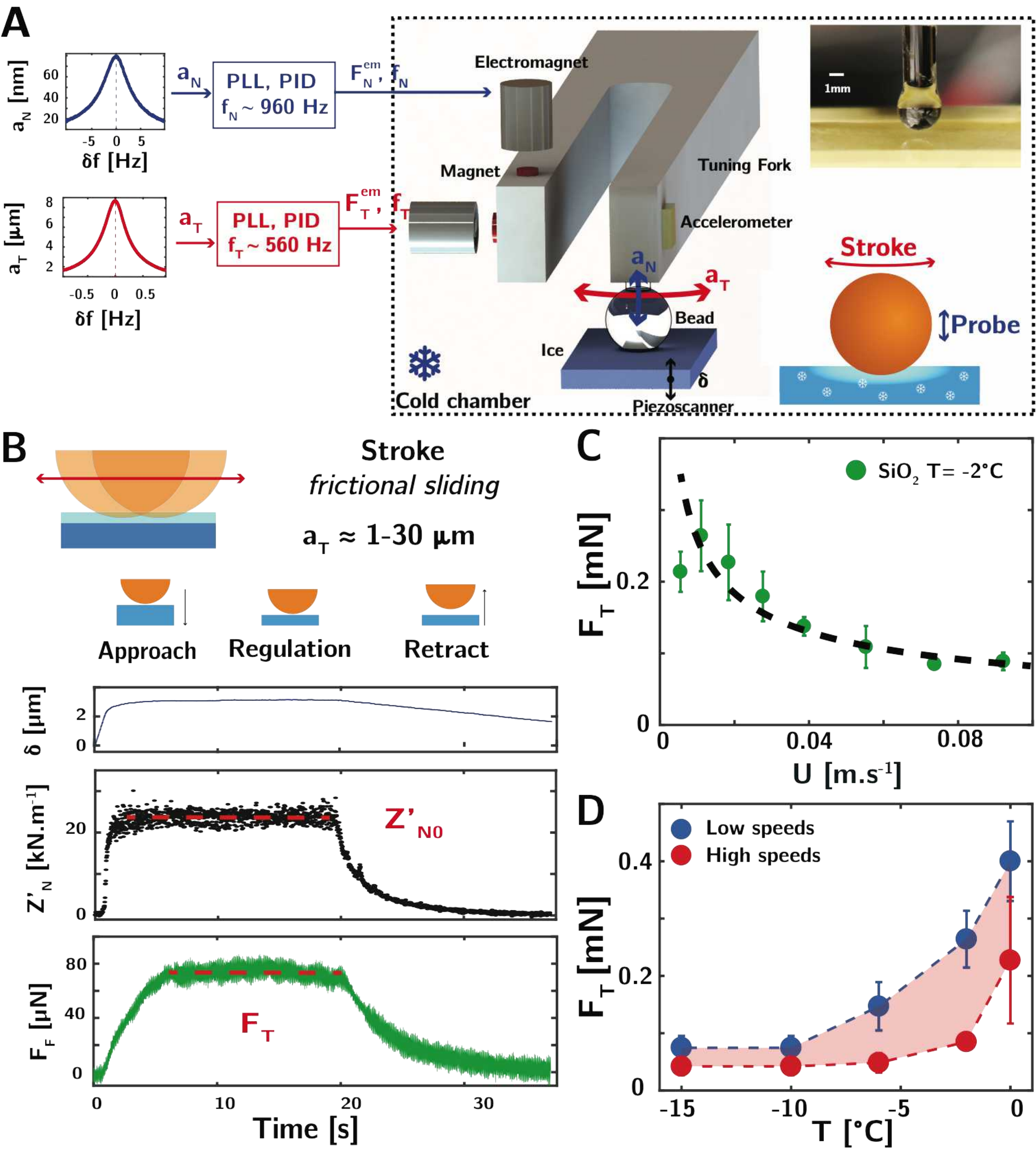}
   \caption{ \textbf{Stroke-probe tribology of ice  (A)} Schematic of the set up. A glass bead of radius $R \approx 1.5$ mm, is glued to a macroscopic tuning fork (not to scale). The tuning fork is excited at its resonances, leading to a tangential and normal bead oscillation in order to simultaneously shear the ice (stroke, horizontal red arrow) and measure the interfacial properties (probe, vertical blue arrow). \textbf{(B)} Typical approach-retract curve for the indentation distance $d$ and the tangential frictional force $F_\text{F}$ ($T=-6$\degree C) at U=0.005m.s$^{-1}$. Regulating at prescribed value $Z'_{N0}$ allows us to define a steady-state friction force $F_\text{T}$. \textbf{(C)} Steady-state friction force $F_\text{T}$ as a function of tangential speed $U$. 
   The dashed line is a fit according to $F_\text{T} \propto U^{-\gamma}$ with $\gamma = 0.5$.
   \textbf{(D)} Friction force $F_\text{T}$ as a function of the ice temperature for two distinct velocities ($U=0.01$ m.s$^{-1}\text{ and }U=0.1$~m.s$^{-1}$), exhibiting a steady increase close to the melting point.} 
    \label{fig1}
\end{figure}

 Simultaneously, we take advantage of the higher order eigenmodes of the tuning fork: as sketched in Fig.~1A, we
excite the first normal mode associated with a resonance frequency $f_\text{N} \simeq 960$~Hz ($K_\text{N} \sim 10^3$~kN.m$^{-1},\ Q_\text{N} \sim 200$) and measure the corresponding normal force using a similar procedure as for the tangential oscillation. This gentle probe, with a tiny normal oscillation amplitude $a_\text{N}\sim 50$ nm,  allows us to measure the normal mechanical impedance of the sheared ice $Z_\text{N}^{*}= {F_\text{N}^{*}}/{a_\text{N}}$ with $F^*_\text{N}$ the complex normal force acting on the sphere. 
The real and imaginary part of $Z_\text{N}^{*}= Z'_\text{N}+i\,Z''_\text{N}$ correspond respectively to the normal conservative and dissipative mechanical impedance of the interfacial medium.  We verified that the normal mode oscillation does not influence the tangential mode and the friction measurement (see Supplemental Materials, Figs. S1, S6 and S11) and that the interfacial mechanical properties do not depend upon the normal oscillation amplitude ( Supplemental Materials Fig.~S7).

Altogether, this superposition methodology allows us to probe gently the mechanical properties of the interface while the tangential stroke mode shears laterally the ice surface, echoing the principle of superposition rheometry \cite{Vermant1997}. \LB{It allows performing simultaneously  tribometry and rheology of the contact. }\LC{We emphasize that the experimental methodology and setup were validated by several control experiments on different fluids and conditions: (i) a standard silicone oil was investigated using our protocole, leading to the expected newtonian rheology with the tabulated viscosity, see Supplemental Material Fig.~S11 and Ref.\cite{Laine2019}; (ii) ionic liquids were investigated using this protocole in a previous study, see Ref. \cite{Laine2019} and, beyond the rheology, also evidenced  molecular layering at the interface, confirming the sensitivity of the setup;} 
\LBB{(iii) the two modes (stroke-probe) methodology was successfully implemented to study the interfacial properties of liquids exhibiting complex (shear-thickening) rheology, beyond that of newtonian fluids, see Ref. \cite{Comtet2017}.}
\LB{Last but not least, experiments with an alternative phase-changing material, namely solidified polyelethylene glycol 1000 (PEG-1000), were also conducted in the present study, highlighting interesting similarities with ice which will be discussed later.}

\par
Let us describe now the experimental procedure.  
Prior to the measurements, we first proceed to several preliminary approach-retracts of the probe under sliding, which allows levelling the surface. 
Then, the probe is slowly put in contact with the ice, see Fig.~1B: 
 the indentation distance $\delta$ increases and the friction force increases as the probe starts shearing the ice.
The typical maximal indentation is kept small, typically $\delta_0\sim 3\,\mu m$, and precludes ploughing dissipation, which has been evidenced at higher loads and indentations \cite{Bonn2018}. 
In the experiments, we set the normal conservative impedance $Z'_N$ at a prescribed value $Z'_{N0}$ by adjusting the maximum indentation position $\delta_0$. This 
amounts to fixing the normal load on the sphere, which is obtained by integrating the gradient $Z'_{N}$ over the indentation distance (see Supplemental Materials Fig.~S4). 
This accordingly allows measuring a lateral frictional force $F_T$ for the prescribed load, see Fig.~1-B (horizontal red dashed line). Afterwards, the bead is retracted slowly and the friction force decreases back smoothly to zero. 
 Note that during the retract the normal conservative impedance does not exhibit a square root dependence with the indentation as it would be expected for a Hertzian deformation; \LB{it rather exhibits a hydrodynamic behavior, pointing to a fluidified interface}. We come back more exhaustively below on the impedance's behavior.
Altogether the  procedure described allows us to get reproducible measurements under a fixed load at different contact locations, see Supplemental Materials Figs.~S3-S4.
\LC{Unless specifically mentioned,  measurements for both frictional and rheological results presented in the main text are performed for the same value $Z'_{N0}=24$ kN.m$^{-1}$, corresponding to a load of 4.5 mN}. The effect of the load is specifically studied in the supplemental materials, section S3.2 \cite{SM}.
\par
\section{Tribology and rheology}

\subsection{Friction}
We first report in Fig.1C the lateral friction force $F_\text{T}$ as a function of the tangential velocity $U$ (associated with $a_\text{T}$ in the range $1-30~\mu$m). The friction force does not vanish at low speeds, similarly to solid-on-solid friction. In addition,
a weak power-law decay of the frictional force versus velocity is observed, with $F_\text{T} \propto U^{-\gamma}$ with $\gamma \sim 0.3-0.5 $. We emphasize that this behavior is consistent with previous macroscopic measurements on ice (and snow) \cite{Buhl2001, Baurle2006} (see also Supplemental Materials Fig.~S5-B). 
%
Furthermore the friction force at a fixed velocity is found to be proportional to the normal load (Supplemental Materials Fig.~S9-A, S5-A). This points to a solid-like friction
characterized by a friction coefficient $\mu=0.015$; this value  is also in very good agreement with macroscopic measurements on ice \cite{Bonn2018}. 
Finally, repeating these measurements for various temperatures allows us to obtain 
the temperature dependence of the friction force, see Fig.~1D. 
Counterintuitively, the friction force (here shown for two different velocities) is shown to increase steadily as the melting point is approached, echoeing similar observations in other macroscopic experiments \cite{Baurle2006, Bonn2018}. 

\begin{figure}[!htb]
\centering
  \includegraphics[width=\columnwidth]{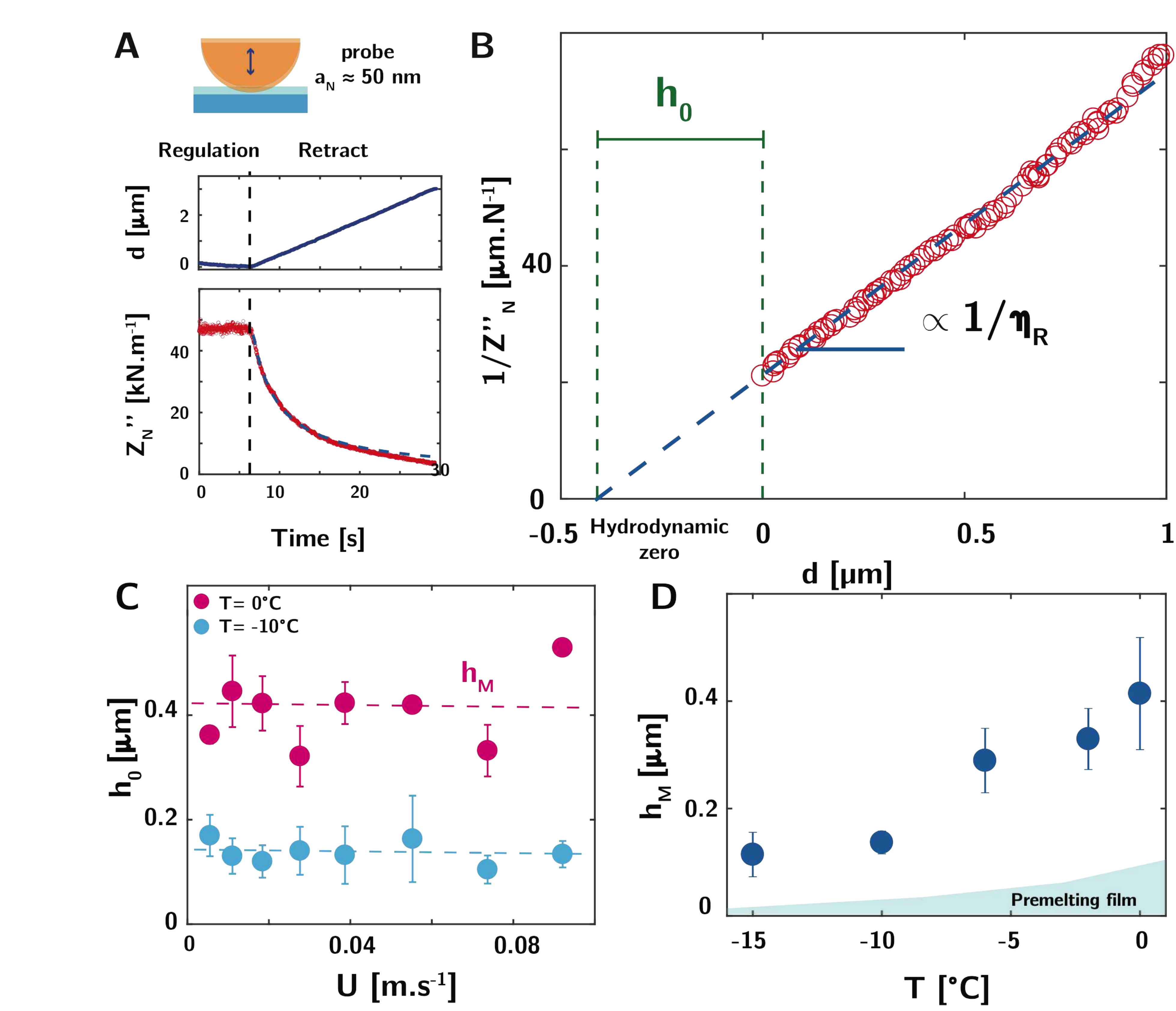}
   \caption{ \textbf{Measuring the thickness of the interfacial film} \textbf{(A)} Retract curve for the dissipative ($Z''_\text{N}$, red) normal mechanical impedance while the bead is simultaneously shearing the ice surface at speed $U=0.01$m.s$^{-1}$ ($T =-2$\degree C). 
   Under regulation, a steady state is reached. Upon withdrawal, the impedance $Z^*$ relaxes smoothly to zero. \textbf{(B)} Inverse of the mechanical impedance $1/Z_\text{N}''$ as a function of the retract distance $d$ (during the retract phase).
The linear variation of $1/Z_\text{N}''$ versus the distance is accounted for by the Reynolds formula, Eq.(1), and points to the liquid-like nature of the interfacial layer. 
  \textbf{(C)} Measured thickness $h_0$ as a function of speed for two temperatures ($T=-10 $ \degree C and $T=0$ \degree C). In each case, there is no variation with the shearing speed U. Measurements are performed under the same load L=4.5mN.
  \textbf{(D)} Average thickness $h_M$ as a function of T. As intuitively expected, we observe an increase in the film thickness as we approach the melting point. An upper limit for the thickness of the premelting film on ice taken from literature is shown in pale blue \cite{Michaelides2017}.} 
   \label{fig2}
\end{figure}

\subsection{Interfacial mechanics}
Now, taking advantage of the normal probe mode, one has access to the mechanical properties of the interfacial film under sliding.
Both the real ($Z'_\text{N}$) and imaginary ($Z''_\text{N}$) parts of the mechanical impedance, which are respectively related to the elastic and dissipative response of the interface (Supplemental Materials, Eqs. S.1, S.2), are deduced. The variations of the normal mechanical impedance $Z_\text{N}''$ under contact and upon retract are shown in Fig.~2A. We observe the same trends as for the friction force $F_f$ (see Fig.~1B): a plateau during the regulation at $Z'_{N0}$ followed by a smooth decrease during the retract. Further insights into the dissipation are obtained by plotting the inverse of the  dissipative impedance $1/Z''_\text{N}$ as a function of the retract distance $d$ see Fig.~2B. A first key result from this plot is that a {\it linear variation} of the inverse normal impedance is measured as a function of the retract distance $d$. Only far from the contact (large $d$) a small deviation from this linear behavior can be observed. This suggests a hydrodynamic-like response of the interface during the withdrawal, consistent with the Reynolds law:
\begin{eqnarray}
Z''_\text{N} =  \frac{6 \pi \eta_{R} R^2  \cdot \omega_\text{N}}{{h_{\rm hyd}}}  
\label{equationReynolds}
\end{eqnarray}
where R is the sphere radius, $h_{\rm hyd}$ is the hydrodynamic film thickness and $\eta_{R}$ the viscosity.  \LC{For a non-vanishing shear velocity $U$, the interstitial fluid exhibits a viscous-like response during the retract. }

\subsection{Interfacial film thickness}
\LB{In the stationary state, the hydrodynamic film thickness $h_{\rm hyd}$ is not fixed {\it a priori} but self-adjusts to reach a stationary value. According to the linear relation between $h_{\rm hyd}$ and $1/Z''_\text{N}$ highlighted above, the thickness $h_0$ of the stationary film can then be obtained from the measurement of the dissipative modulus $Z''_\text{N}$. As shown in Fig.~2A in the regulation regime,   $Z''_\text{N}$ reaches a plateau as a function of time under imposed shear velocity $U$ and normal load. Accordingly the stationary film thickness $h_0$ is deduced from this plateau value thanks to Eq.(\ref{equationReynolds}). 
Equivalently, $h_0$ can be read on the graph in Fig.~2B by extrapolating the   $1/Z''_\text{N}$ line versus  $h_{\rm hyd}$ to the zero intercept (dashed blue line):
$h_0$ then corresponds directly to the absolute value of the extrapolated hydrodynamic zero (Fig.~2B, green dashed line).}
\LC{Note that by convention, and to make the reading easier, we set in Fig.2-B the position of the sphere corresponding to the plateau value to $d=0$, so that $h_0$ corresponds directly to the absolute value of the extrapolated hydrodynamic zero.  
In principle the hydrodynamic thickness is expected to be the sum of the actual film thickness and a slip length, if any. However since ice is hydrophilic, a very small slip length is expected (typically few nanometers) \cite{Bocquet2010}, so that the hydrodynamic thickness should be safely identified to the real film thickness. }

Repeating this procedure under various experimental conditions enables us to retrieve the stationary film thickness as a function of the lateral sliding speed, normal load and temperature in the considered sphere-plane geometry.
We report in Fig.~2C, the variation of the interfacial film thickness $h_0$ with the tangential speed $U$. Surprisingly, $h_0$  is found to barely depend on the tangential speed, in contrast to the common belief that a larger velocity would induce a larger film \cite{Oosterkamp2019,Colbeck1988}. Similarly its shows a weak variation as a function of the normal load (Supplemental Materials Fig.~S9-C). However, the thickness increases with temperature (Fig.~2D)  ranging from 100 nm to 500 nm. \LB{To fix ideas,} \LC{the thickness of the stationary film is typically a factor of four higher than the values for the equilibrium premelting films (pale blue in Fig~2.D)  \cite{Doppenschmidt2000, Kornienko2013,Wettlaufer1999, Conde2008}.} 

\begin{figure}[!htb]
\centering
  \includegraphics[width=\columnwidth]{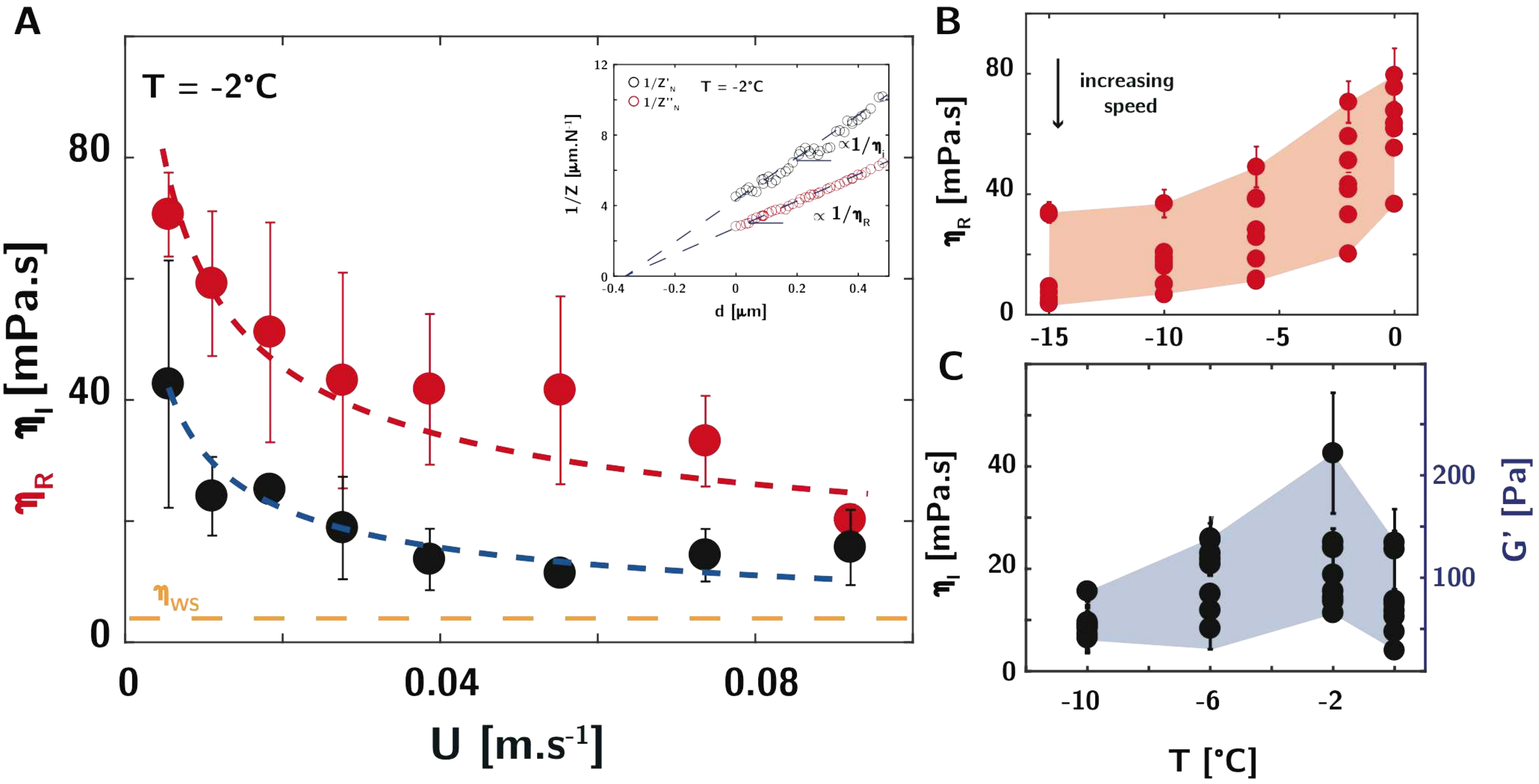}
   \caption{\textbf{A visco-elastic interstitial film} \textbf{(A)} Evolution of the real and imaginary part of the viscosity as a function of the shearing speed $U$. The dashed lines correspond to a fit $\eta_{R,I} \propto U^{-\alpha}$ with $\alpha=0.5$. \textbf{(B)} Evolution of the real part of the viscosity (dissipation) as a function of temperature: we observe a steady increase towards the melting point, reminiscent of the friction force trend (see Fig.~1C).  \textbf{(C)} Imaginary (elastic) part of the viscosity of the interstitial fluid as a function of temperature. The right axis of panel (C) provides the corresponding elastic modulus $G'$.
    }
   \label{fig3}
\end{figure}

\subsection{Rheological properties under shear}
A second fundamental  lesson emerging from these measurements is that the interfacial film under shear exhibits a visco-elastic rheology, associated with a complex viscosity $\tilde \eta  = \eta_\text{R} - i \eta_\text{I}$ \cite{Larson1999,Cross2018}. Indeed, as shown in the inset of Fig.~3A and in Fig.~S8, the inverse of the elastic impedance $1/Z'_{N}$ also exhibits a linear variation with the separation distance $d$ during the retract, allowing to retrieve both real and imaginary parts of the viscosity, $\eta_R$ and $\eta_I$, from the corresponding slopes. In the experiments under various conditions, we observe that the linear extrapolations of the \LC{inverse} elastic and   dissipative moduli cross at the same hydrodynamic zero within 30\% error, see Fig.~S8. Such visco-elastic response of the interstitial film is analogous to that of complex fluids, {\it e.g.} polymers and polyelectrolytes \cite{Larson1999,Cross2018}.
Quantitatively, a first striking result is that the \LB{measured viscosity $\eta_\text{R}$ under shear} is much higher than the typical viscosity of supercooled water at the same temperature 
$\eta_{\rm W,S}$, see Fig.~3A (orange dashed line)  \cite{Hallett1963}. Both the real and the imaginary part of the viscosity follow a weak power law decay as a function of the tangential speed, similarly to the friction force: $\eta_{R,I} \propto U^{-\alpha}$ with $\alpha \sim 0.3-0.5 $, see Fig.~3A. 
Finally, $\eta_\text{R}$ is found to increase tremendously towards the melting point and reaches a value close to two orders of magnitude higher than water at 0\degree C (Fig.~3B). 
On the other hand, the imaginary (elastic) part $\eta_I$ is found to be less sensitive to temperature, but the corresponding elastic modulus $G'$, {$G' =  2 \pi f_\text{N}\,\eta_\text{I}$} typically lies in the range of $10^2$ Pa (Fig.~3C), stressing the strong elastic response of the film. 
As a side note, one may remark that the measured values for the elastic impedance $1/Z'_{N}$ in the stationary state is much larger
than estimates for a capillary contribution due to a meniscus : $\gamma R/h_0 \sim 1$kN.m$^{-1}\ll Z'_{N0}\sim 20$kN.m$^{-1}$ with $\gamma$ a typical surface tension, and furthermore with opposite sign. 
Moreover the estimated dissipation due to the contact line displacement and meniscus dissipation would be in the order of few $\mu$N, almost two orders of magnitude smaller than the measured friction force. This shows that capillary effects are negligeable here.

\LC{Altogether, our observations converge to an unexpected complex rheology for the meltwater. A first comment 
is that the interfacial water film under shear is `as viscous as an oil', with a viscosity up to two orders of magnitude larger than bare water. 
This points to an unexpected rationale for the exceptional friction properties of ice, contrasting with the bad lubricant behavior of bare liquid water. 
Indeed a viscous film is a prerequisite to properly lubricate the contact: it limits squeeze-out, thereby avoiding direct solid-on-solid contact. 
In contrast to standard water, the 'slimy melt water' which is generated under sliding is an excellent lubricant. 
The complex rheology of meltwater 
has been completely overlooked up to now in the modeling of ice friction. The latter usually assumes bare newtonian water and focuses on the interplay between frictional heating and the thickness of the meltwater film. Our findings suggest to reconsider the standard framework 
for ice friction, as well as the dissipative mechanisms occuring in the lubricating film.}

\subsection{Effect of hydrophobic coatings}
\LC{Last but not least}, a puzzling standard practice in winter sports is to use hydrophobic coatings to reduce friction, typically wax containing fluor additives \cite{Bowden1955}. However, adding hydrophobic coatings to favor water lubrication may seem counterintuitive and the very origin of this behavior remains mysterious. To this end, we have investigated the friction properties of hydrophobic silanized  silica beads. Here the sliders differ from the previous glass spheres only by a nanometric silane layer coated on the surface of the bead.
\begin{figure}[!htb]
\centering
  \includegraphics[width=\columnwidth]{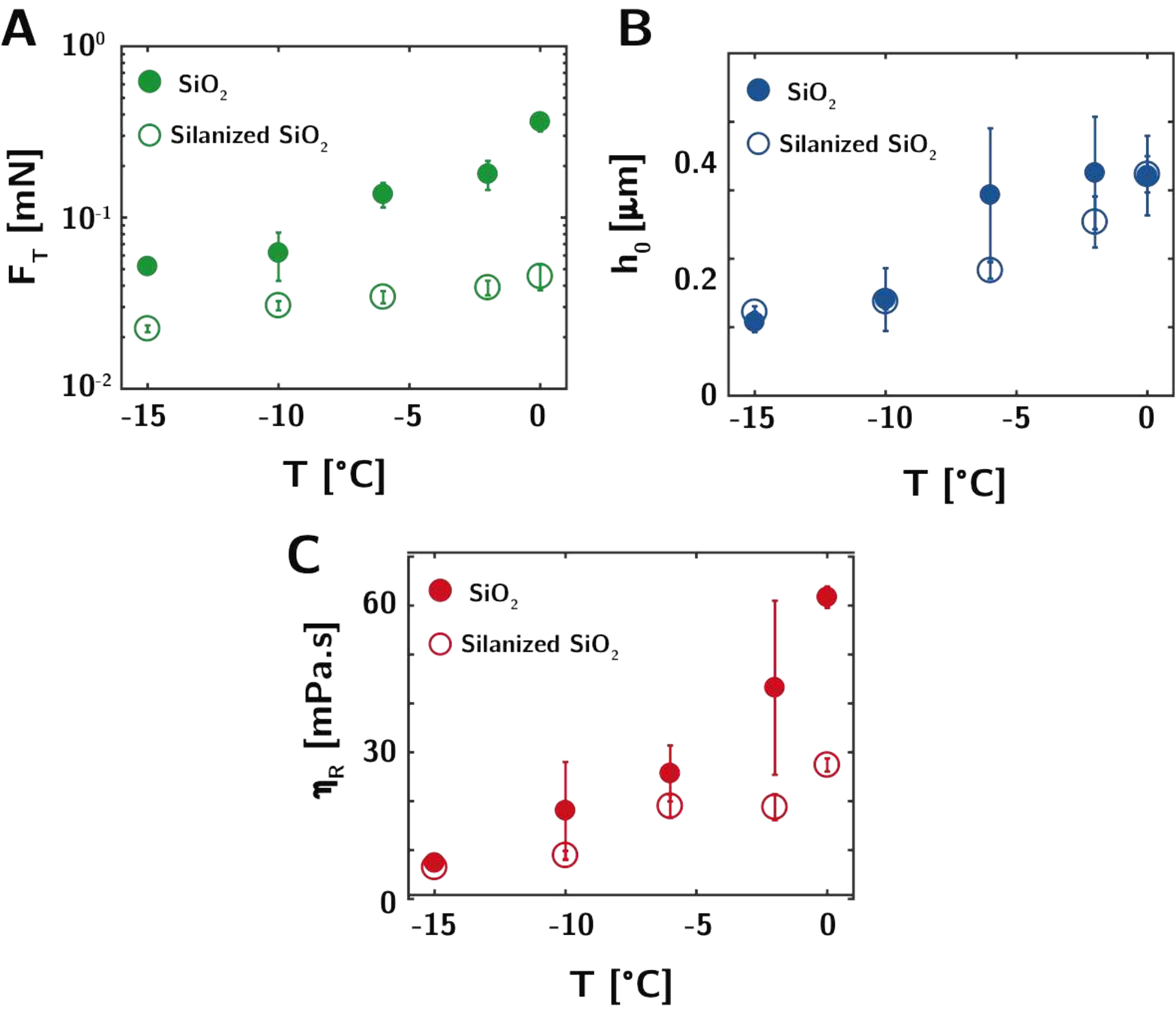}
   \caption{\textbf{Effects of the hydrophobic coatings} \textbf{(A)}  Comparison of the friction force for a silanized sphere and bare glass for $U=2.8$cm.s$^{-1}$ as a function of temperature. We measure a drastic reduction of friction for the hydrophobic glass sphere.  \textbf{(B)} Comparison of the film thicknesses between the hydrophilic and hydrophobic coatings, showing similar results. \textbf{(C)} Comparison of the dissipative part of the viscosity: at high temperatures, the hydrophobic coating highlights a lower viscosity $\eta_{R}$, in line with the observed reduced friction. The effect is reduced at lower temperatures. 
   }
    \label{fig5}
\end{figure}
 As highlighted in Fig.~4A, the hydrophobic treatment leads to a drastic reduction of friction, as much as a factor of 10, as compared to standard glass surface. The friction reduction becomes stronger close to the melting point. This is in agreement with observations for the effect of wax on snow friction \cite{Bowden1955}. To get more insights into the dissipative mechanism at stake, we plot in Fig.~4B and 4C the respective film thickness and viscosity $\eta_R$ for the hydrophobic and hydrophilic sliders. Our measurements  highlight that this friction reduction is not associated with a modified hydrodynamic film thickness $h_{0}$ (Fig.~4B). On the one hand, this is somehow surprising since it excludes {\it a priori} the effect of a finite hydrodynamic slippage at the surface, which may occur for complex fluids \cite{Bodiguel2013,Cross2018} \LB{(slippage would indeed enhance the hydrodynamic thickness of the film according to $h_0=h_{\rm layer}+b$, with $b$ the slip length, which is not observed)}. On the other hand, it confirms that the hydrodynamic thickness that we measure is not affected by a slippage effect, even in the case of bare glass. 
  Rather, we observe a clear reduction in the real part of the viscosity for the hydrophobic glass as compared to the hydrophilic case. This effect is amplified as the temperature approaches the melting point (Fig.~4C). The trend is also qualitatively the same as for the friction force. The relationship between the viscosity of the interstitial medium and the surface properties suggests that hydrophobicity may affect \LC{the build up of the interstitial film.} 
Altogether, these findings show that surface effects {\it at the nanometric scale} can strongly impact macroscopic ice friction. 
  We note that in winter sports, wax coatings are not only  hydrophobic, but their composition are chosen such that its hardness adjusts the ski sole to that of the snow/ice grains. Actually, snow is a much more complex material than ice: it is a porous material involving a mixture of soft snow, hard ice and water. While our results shed some light on the hydrophobic effects, the specific interplay beween elasticity, {wear} and hydrodynamic properties at stake in snow has still to be completely uncovered.

\subsection{Alternative material}
\LC{Beyond ice, it is interesting to compare these results with an alternative material, which might also undergo a phase change under shear. To this end, 
we performed a similar study 
on solid polyethylene glycol 1000 (PEG 1000).} 
\LB{This waxy material is solid at room temperature, but becomes liquid around 35\degree C.}
\LB{We followed the very same experimental procedure as detailed above for ice, exploring its frictional and rheological response under shear at room temperature ($\sim 24$\degree C). The results are reported in the Supplemental Materials, Figs.~S12 and S13.} \LC{The lateral friction force $F_T$ is measured to be an affine function of the shear velocity, $F_T=F_T^0 + \alpha\, U$,  with a finite friction force $F_T^0\simeq 0.1$ mN as $U\rightarrow $0 and $\alpha\simeq 0.015$ kg.s$^{-1}$ a friction coefficient}.
\LB{In parallel, the measurements of the rheological properties of the interface using the normal mode highlight a viscous-like response in the tested range of shear velocities.}
%
\LC{The viscosity is measured to be $\eta_R \simeq 0.5$Pa.s, and independent of $U$. Notably, this value is of the same order (but slightly larger) as compared to the viscosity of PEG1000 at 35\degree C -- the fusion temperature --, which we measured by standard rheometry, see Supplemental Materials Fig.~S14.}
\section{Summary and Discussion}
\LB{To summarize, 
these results reveal interesting similarities between ice and PEG 1000. In both cases, a finite friction force is measured in the limit U$\rightarrow 0$, {\it i.e.} a yielding behavior of the interstitial film with a threshold force (or stress) to induce flow. Furthermore, the rheological response of the interstitial film under lateral shear exhibits a hydrodynamic-like response in the tested velocity range. }
\LC{Overall the film response is  intermediate between that of a pure solid and a pure liquid, with a yielding behavior and shear-induced fluidization.}
This behavior bares  similarities with solid-on-solid friction where the interstitial joint is shown to exhibit the  phenomenology of soft glasses \cite{bureau2002rheological}. 
\LB{In a  different context, this also echoes closely the observation for fluidization of granular materials which is induced by an independent flow agitation and result in an effective viscosity of the (otherwise yielding) material \cite{vanHecke2012}}.
\LC{The emerging picture is in contrast with the prevailing theories for ice friction, which assume a straight transition from a cristalline phase to the bare liquid water phase.
\LB{However it is consistent with the fact} that shear can not lead to a full liquefaction of the contact since this would require a very strong temperature increase in the contact, in contrast with various experimental observations in our experiments, see Supplemental Materials Fig.~S10, as well as others Refs.\cite{Baurle2006, Lever2018}.}
\LB{The recent experiments of Ref. \cite{Lever2018}, reporting microscale infra-red thermography and optical measurements of snow-grain contacts, actually highlighted abrasion, rather than melting, of the interstitial contact region.} 
Gathering these experimental observations, a tempting explanation for the observed  response is accordingly that, under abrasive wear, a suspension of liquid  and sub-micron (ice or PEG) debris is formed, \LB{hence resulting in composite lubrification in the contact}. \LB{Lateral shear is expected to fluidize the \{solid-liquid\} mixture within the interstitial layer, again similar to \cite{vanHecke2012}, and the interstitial material should exhibit a complex rheology typical of dense suspensions \cite{Boyer2011}. The only difference is that here the grains constituting the suspensions should self-adjust under the imposed shear, load and temperature. }    For ice, the increase of viscosity with temperature may be interpreted as an increasing density of ice fragments when the ice becomes softer close to the melting point; also higher normal loads may also lead to higher indentations and abrasions, providing more debris and higher viscosities. As a matter of fact, the viscosity extrapolated at vanishing load, obtained  from the linear variation of the viscosity with the load (Supplemental Materials Fig.~S9-D), does perfectly match the viscosity of supercooled water at -6\degree C, suggesting again that the further rise is a direct consequence of the abrasion during the contact. We leave for future works the full structural characterization of the interstitial material under shear, {\it e.g.} using spectroscopic measurements.

 
Overall, our results thus call for a deep overhaul of prevailing theories of ice friction. The complex rheology of the interstitial material is shown to be a key ingredient, which has not been considered up to now in the theoretical frameworks describing water as a newtonian fluid. Modelling the intertwinned relationships between the mechanical, rheological and thermodynamic mechanisms is challenging but our results provide a  guide and an experimental benchmark to revisit the standard framework of ice friction. 

\section{Conclusion}
 Thanks to our unique stroke-probe experimental setup, we have been able for the first time to bridge the gap between nanoscale and macroscale tribometry of ice, while fully characterizing the mechanics of the ice interface during frictional sliding. A key outcome of our study is the evidence for a complex mechanical behavior of the interstitial meltwater, \LB{which exhibits the rheology of a complex yielding  material}. 
 Its large viscosity, coupled to an elastic response, yields an excellent hydrodynamic lubricant behavior, leading {\it in fine} to low friction. Our experiments challenge the existing theories  and should motivate a complete reformulation of  the frameworks describing ice friction on the basis of the new fundamental insights unveiled here. 
  Finally the self-lubricating behavior of ice suggests to develop soft and phase-changing solids as antiwear tribofilms \cite{Carpick2015}. 
  Ionic liquids, which have been shown to exhibit freezing in metallic confinement \cite{Comtet2017b}, are good candidates in this perspective.


\subsection*{Acknowledgments}
The authors are indebted to Martin Fourcade,  Gr\'egoire Deschamps, Yann Guigonnet and Baptiste Claudon from the french biathlon team for many fruitful discussions on ski and waxing. {The authors thank Axel Laborieux and Adrien Benusiglio for contributions at the early stages of the project, Joshua McGraw for his help with the silanization process and Quentin Louis for his contribution to the experiments on polyethylene glycol. We also thank Annie Colin for
discussions and help on the rheometry measurements.
LB is indebted to Ludwik Leibler and Elisabeth Charlaix for many interesting discussions. The authors also thank {Thomas Salez, Liliane L\'eger} for their insights and feedbacks. 
We acknowledge support from ANR grant {\it Neptune}. 
AS acknowledges funding
from the European Union's Horizon 2020 Framework Program/European
Research Council Starting Grant 637748 NanoSOFT.  LB acknowledges support
from the European Union's Horizon 2020 Framework Program/European
Research Council Advanced Grant 785911 Shadoks.}



\end{document}